# Why Cheating is Wrong

Scott Williams and Michael Courtney
U.S. Air Force Academy, 2354 Fairchild Drive, USAF Academy, CO, 80140-6210
Michael.Courtney@usafa.edu

**Abstract:** Mathieu Bouville's "Why is cheating wrong?" (Studies in Philosophy and Education, 29(1), 67-76, 2010) misses the mark by failing to consider the longer term consequences of cheating on student character development and longer term societal consequences of undermining professional expertise and trust in disciplines where an earned degree is an essential part of professional certification and qualifications. Educators who turn a blind eye to student cheating are cheating the public by failing to deliver on the promise of graduates who genuinely earned their degrees.

Intellectual gymnastics (Bouville 2010) do not change the fact that cheating is wrong. In a series of self-limiting arguments, the author repeatedly dismisses the negative effects of cheating, suggests cheating is essentially equivalent to dysfunctional pedagogy, and claims cheating is therefore wrong only to the extent that it has material consequences on learning and assessment. That cheating – to practice fraud or deceit (cheating n.d.) – is wrong independent of academic consequences is dismissed. Yet the objective wrongness of cheating is the central issue. Moreover, while the author offers legitimate criticism of common pedagogical practices, the uses and efficacy of grades, and the sometimes misplaced focus of the academic system, his attempts to dismiss the effects of cheating in light of these concerns ring hollow. Imperfections in sincere efforts to engage and assess student learning cannot be equated with deliberate attempts to defraud the system.

The negative effects of cheating go far beyond immediate issues such as diluting the meaning of grades, creating inequity between students, or undermining the learning environment, even if the effects of these things are less substantial than is generally perceived. Since actions form habits, and habits form character, academic dishonesty builds into the character a propensity for dishonesty. In addition, since academic credentials are criteria for professional certifications, academic dishonesty carries the risk of the unfounded illusion of professional competence. How many readers relish the thought of being treated by doctors who cheated in their anatomy and physiology courses, or having important lab tests performed by technicians who fraudulently secured (via cheating) their required certifications?

An individual's character and integrity are paramount. Here at the United States Air Force Academy, our mission is to commission leaders of character who are prepared to honorably serve their nation. The nation expects and requires its military officers to uphold their oaths of office, adhere to the highest standards of conduct, effectively lead those in their command, and steward both weapons of war and secrets of national security. Furthermore, earned degrees are required components for certification of professional competence in many areas of military service. Cheating at any stage of officer development and regardless of immediate consequences is therefore absolutely incompatible with the profession of arms. What nation wants military officers in charge of navigation, engineering, or other technical issues related to national security who cheated their way through coursework rather than demonstrating genuine competence in required subject areas? What nation is eager to entrust the lives of its sons and daughters or its weapons of war to those who cannot even demonstrate faithfulness in college coursework?

It is sophistry to argue, "Breaking a rule is illegitimate only if the rule is legitimate. Either the rule has a rational justification and this rather than breaking a rule makes cheating wrong, or the rule is arbitrary and there is no reason to endorse it." (Bouville 2010) The legitimacy of rules rests in the legitimacy of the issuing authority; it is not given to individual students to whimsically decide whether or not the rules apply to them, especially when the student has agreed (implicitly or explicitly) to the rules by virtue of enrollment. To claim otherwise is to promote anarchy – a chaotic system with no real standards as students choose for themselves what feels right to them. How many citizens are eager to live in a country where the police and the military only follow the rules they deem to have adequate rational justification? The human mind has infinite capacity for finding flaws in the "rational justification" of rules that the human heart is inclined to disobey. Due process in the making and enforcing of rules and laws of orderly society does not require "rational justification" to the satisfaction of every individual who has a duty of compliance.

People expect their doctors, their pilots, their engineers, and their military officers to have genuinely earned their professional credentials and to meet rigorous standards in areas of knowledge and conduct necessary for public trust in the performance of their duties. Cheating is wrong because academic dishonesty in the training of these professions undermines both the expected level of expertise and the expected level of trust. Educators have a duty to society to ensure the quality of graduates, and this duty includes good faith efforts to prevent academic dishonesty.